\RequirePackage{fix-cm}
\documentclass[smallcondensed]{svjour3}     
\smartqed  

\usepackage{mathptmx}

\usepackage{hyperref}
\usepackage{graphicx}
\usepackage{epstopdf}

\begin{document}

\title{Statistical analysis of sizes and shapes of virus capsids and their resulting elastic properties}
\titlerunning{Statistical analysis of sizes and shapes of virus capsids and their resulting elastic properties}

\author{
An\v{z}e {Lo\v{s}dorfer Bo\v{z}i\v{c}}
\and Antonio {\v{S}iber}
\and Rudolf {Podgornik}
}

\institute{
A. {Lo\v{s}dorfer Bo\v{z}i\v{c}} \and R. {Podgornik} \at
Department of Theoretical Physics, Jo\v{z}ef Stefan Institute, SI-1000 Ljubljana, Slovenia\\
\email{anze.bozic@ijs.si}
\and
A. {\v Siber} \at
Institute of Physics, 10001 Zagreb, Croatia\\
\and
R. {Podgornik} \at
Department of Physics, Faculty of Mathematics and Physics, SI-1000 Ljubljana, Slovenia
}

\date{Received: date / Accepted: date}

\maketitle

\begin{abstract}

From the analysis of sizes of approximately 130 small icosahedral viruses we find that there is a typical structural capsid protein, having a mean diameter of 5 nm and a mean thickness of 3 nm, with more than two thirds of the analyzed capsid proteins having thicknesses between 2 nm and 4 nm. To investigate whether, in addition to the fairly conserved geometry, capsid proteins show similarities in the way they interact with one another, we examined the shapes of the capsids in detail. We classified them numerically according to their similarity to sphere and icosahedron and an interpolating set of shapes in between, all of them obtained from the theory of elasticity of shells. In order to make a unique and straightforward connection between an idealized, numerically calculated shape of an elastic shell and a capsid, we devised a special shape fitting procedure, the outcome of which is the idealized elastic shape fitting the capsid best. Using such a procedure we performed statistical analysis of a series of virus shapes and we found similarities between the capsid elastic properties of even very different viruses. As we explain in the paper, there are both structural and functional reasons for the convergence of protein sizes and capsid elastic properties. Our work presents a specific quantitative scheme to estimate relatedness between different proteins based on the details of the (quaternary) shape they form (capsid). As such, it may provide an information complementary to the one obtained from the studies of other types of protein similarity, such as the overall composition of structural elements, topology of the folded protein backbone, and sequence similarity.
\keywords{ capsid; virus; geometry; icosahedron; faceting; elasticity}
\end{abstract}

\section{Introduction}

Viruses are the most abundant source of DNA and proteins in Earth's oceans that contain on the order of $10^{30}$ virions~\cite{Marine}. Yet their status as living entities is often called into question as they do not conform to the ``self-reproduction with variations'' standard~\cite{Trifonov}. It appears that much of their features can be understood in terms of thermodynamic equilibrium physics~\cite{Bruinsma,PCCP,Zandi2004}, especially when they are ``dormant'', i.e. outside the cells which they infect, where they in fact turn into little more than very complicated macromolecules with a ``life cycle''~\cite{Macromol}. The fact that many features of their life cycle can be understood within the equilibrium framework sets them apart from the rest of the living biological systems.

The lack of self-reproduction with variations that we associate with (present-day) life could also mean that viruses predate precellular life~\cite{Koonin2005}, which is partially corroborated by the fact that there seems to be no living system that is immune to viruses, including viruses themselves \cite{Infection}. The long lasting debate about the position of viruses in the great divide between the living and nonliving has been only intensified in recent years with the discovery of viruses with huge genomes that encode proteins which allow for mechanisms that we do associate with life~\cite{LaScola2006,Claverie2010,Arslan2011}. These gigantic viruses are more complex than some bacteria which even further obscures the question of their status and their origin.

The usual approach to trace the origins of a virus is to analyze the virus genome and compare it to other strains and viruses with the hope of uncovering evolutionary relatedness \cite{Holmes}. This can, however, be a futile endeavor as viruses mutate quickly, especially when their genome is RNA based~\cite{Koonin1998}. The phylogenetic approach thus often reduces to comparison of highly divergent sequences, which in effect prohibits the accurate extraction of the evolutionary information and the determination of the relatedness of different viruses. Viruses are thus typically classified only in terms of the type of genome that they contain (double-stranded (ds) or single-stranded (ss) DNA, dsRNA, plus- and minus-sense ssRNA, and reverse transcribing viruses) or whether they wrap themselves in a piece of cellular membrane or not (enveloped and non-enveloped viruses).

The virus phenotypes, on the other hand, are often strikingly similar. A most obvious similarity is between the protein shells (capsids) they form to protect and pack their genomes. A large number of viruses have an icosahedral capsid that consists of many copies of a single or a few very similar proteins arranged in a highly symmetrical manner that can be defined in precise mathematical terms~\footnote{This is the basis of the Caspar-Klug classification scheme~\cite{CK} that classifies different capsid icosahedra in terms of the triangulation number $T$, which is, roughly, a way to divide the icosahedron in similar, nearly equivalent parts that represent individual proteins.}. Thus, even for viruses which are highly diverged and apparently unrelated (e.g. some may infect plant and other animal cells), {\em there are conserved features in their phenotype}, i.e. in the proteins. A viable mode of investigation of virus evolution and origin may thus be through the analysis of capsid protein structure and function, which appear to be fairly conserved, even when the sequences coding for the proteins are very divergent. 

The icosahedral nature of the capsid requires proteins to assemble in precise relations to their neighbors, and disruption of some key aspects of protein structure, such as spatial distribution of hydrophobic/hydrophilic patches and/or charge, may completely block the assembly of a virus from its constituents. This is an evolutionary dead end for a virus, so there are obviously some {\em physical constraints} obeyed and encoded in the viral genome, which must be preserved in the evolution of a virus. The aim of this paper is to accentuate this information through the analysis of shape and size distribution of different icosahedral viruses.

As obvious as it may seem, this is by no means a straightforward and clearly defined task. After all, what does it actually mean to analyze the similarities between the {\em shapes} of different viruses and what sort of an information one gains in the process? We shall seek the answer to this question in the nonlinear theory of elastic shells~\cite{LMN} by carefully comparing a large amount of different viral shape information to numerical predictions, and then extracting a sequence of (F\" oppl-von K\' arm\'an, FvK) numbers pertaining to the sequence of shapes. When dealing with real viruses, one is confronted with an experimentally determined structure that contains spatial coordinates of (ideally) all the atoms that compose the virus. This is a huge amount of data containing a detailed description of the virus surface. Some of these details are of course important for virus attachment (receptor geometry), but there are also some generalities that are expected to be a consequence of the elasticity of the protein shell. Our work presents the attempt to identify the features of the shape related to elastic properties of protein-protein interactions in the capsid, and to numerically quantify these properties.

\section{Analysis}
\label{sec:ana}

\subsection{Structural Dataset Used}

In our analysis we have used approximately 130 capsid entries deposited in VIPERdb~\cite{VIPER}, obtained from X-ray scattering experiments or cryo-electron microscopy. The capsid triangulation numbers range from $T=1$ to $T=p25$~\footnote{The triangulation numbers denoted with $p$ (pseudo) do not completely conform to the Caspar-Klug principle of quasi-equivalence since the basic unit is composed of different (but morphologically similar) proteins.}. The number of viruses with triangulation number $T>1$, used in determining the elastic parameters, is approximately 100, with 29 of them having a $T$-number greater than $T=3$.

To the polyomaviruses and papillomaviruses we assign a triangulation number of $T=6$ instead of the usually used $T=7$ as they are composed of 360 copies of a protein, and should likely be considered as dodecahedral, not icosahedral structures~\cite{Konevtsova2012}. We do this in order for the number of proteins in a capsid to have a clear correspondence with its triangulation number, enabling us a more consistent analysis.

\subsection{Calculation of Best-fitting Prototype Shape and Effective FvK Number of a Capsid}
\label{sec:rmsd}

We have approached the task of quantifying a capsid shape in the following way. The capsid is compared against a set of ``prototype'' shapes whose elastic properties are known (determined). These prototype shapes are introduced and described in Sec.~\ref{sec:fvk}. Of all the prototype shapes we need to pick one which represents the capsid structure, and thus its elasticity, the best. This task requires a definition of a ``quality measure'', i.e. a quantitative parameter describing the quality of the match between the two shapes. The comparison is done by determining the root-mean-square-deviation (RMSD) of the two shapes.

\begin{figure}[!ht]
\begin{center}
\includegraphics[width=7.5cm]{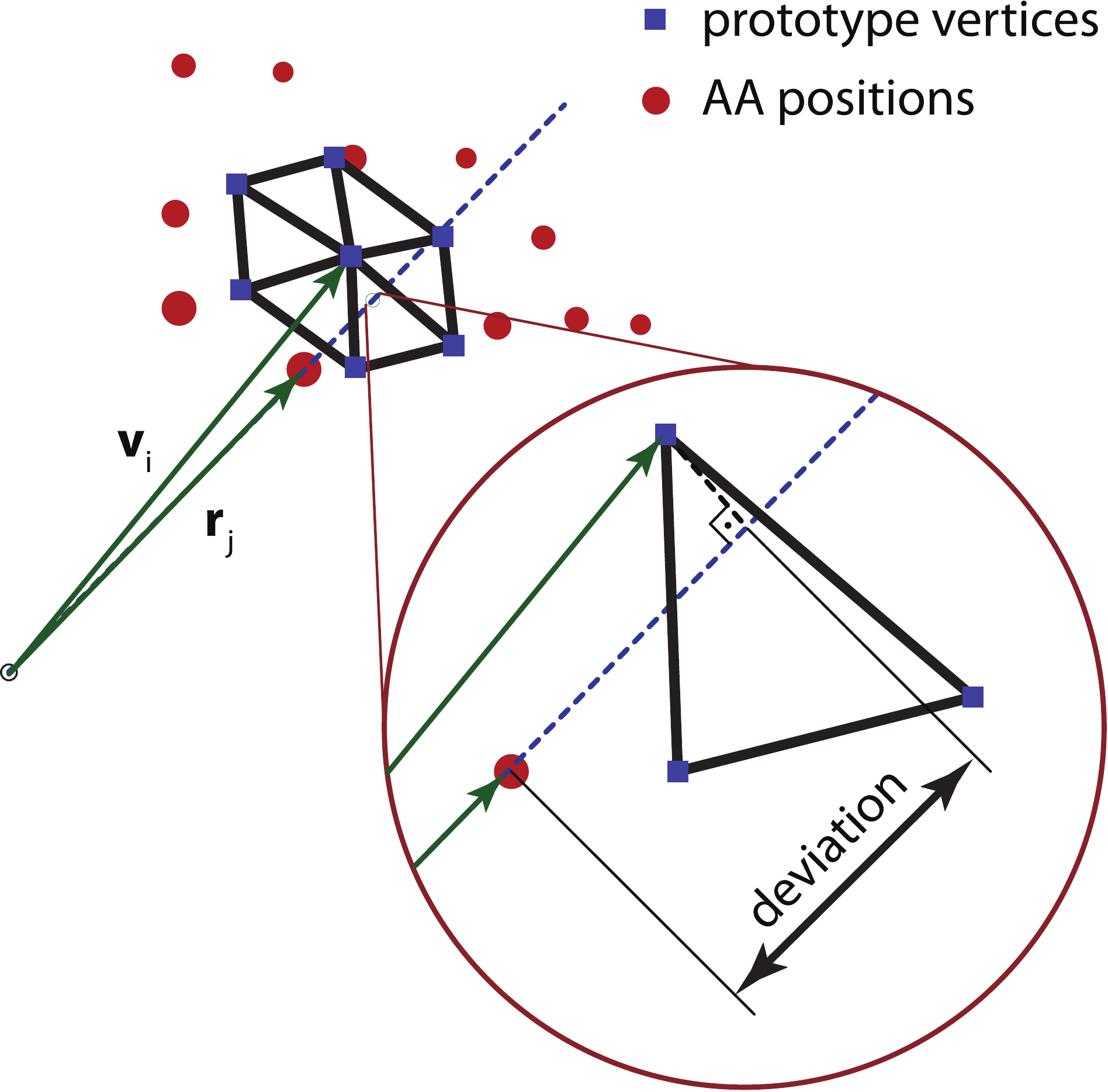}
\end{center}
\caption{A sketch illustrating the pairing of an amino acid $\mathbf{r}_j$ to its closest point on the prototype surface $\mathbf{v}_i^j$. The closest vertex of the prototype $\mathbf{v}_i$ to the amino acid position is determined by the shortest distance from the radial line given by $\mathbf{r}_j$, and the perpendicular distance to the amino acid position then gives the approximation for the deviation $||\mathbf{r}_j-\mathbf{v}_{i}^j||$ used in calculating the RMSD [Eq.~(\ref{eq:rmsd})].}
\label{fig:scheme}
\end{figure}

To calculate the RMSD, we first rotate the prototype shape so that the five-fold symmetry axes of both shapes coincide. For a rotated prototype shape given by the vertices $\mathbf{v}_i$ and the amino acid positions of the capsid $\mathbf{r}_j$ we calculate the RMSD as
\begin{equation}
\label{eq:rmsd}\mathrm{RMSD}(\mathbf{v},\mathbf{r})=\sqrt{\frac{1}{N}\sum_{j=1}^N||\mathbf{r}_j-\mathbf{v}_{i}^j||^2},
\end{equation}
where $\mathbf{v}_i^j$ is the approximation for the closest point on the triangulated prototype surface to the amino acid position $\mathbf{r}_j$~\footnote{The positions of the amino acids are represented either by the positions of their centers-of-mass (see also Ref.~\cite{electro_jobp}), or where this data was unavailable by the positions of their C$_\alpha$ atoms.}. The closest point on the prototype surface to a given amino acid will lie on the radial vector connecting the origin and the amino acid, piercing the prototype surface. Since we do not have the exact surface shape but only its triangulated mesh~\cite{Siber_LMN}, we take for the point $\mathbf{v}_i^j$ the point where the radial vector pierces the tangential plane of the prototype vertex $\mathbf{v}_i$ nearest to the amino acid $\mathbf{r}_j$ (an illustration is shown in Fig.~\ref{fig:scheme}). This gives a good approximation to the true distance from the surface, provided the meshing is fine enough. We have also tried several different distance measures, and found they influence the values of RMSD (a worse approximation yielding bigger values), but not the position of the minimum FvK number for a given capsid. In order to compare the two shapes, the prototype shape and the experimental shell also have to be properly scaled. We thus let the mean radius of one shell to vary slightly during the minimization to allow for a better fit. To minimize the computational time we do the calculation for $1/60\mathrm{th}$ of a virus, since the rest of the amino acid positions in the capsid can be generated by applying the rotational matrices of the icosahedral symmetry group, and thus do not influence the final result.

\section{Results and Discussion}

Before proceeding to the {\em shape} analysis of a virus, it is important for our goal to perform a simple analysis of the average virus {\em size}. This is an information pertinent to the analysis of elastic properties of the capsid, especially the thickness of proteins compared to the average capsid size, as will be shown further on in the paper.

\subsection{Effective Size of Virus Proteins}

\begin{figure}[!htb]
\begin{center}
\includegraphics[width=8.5cm]{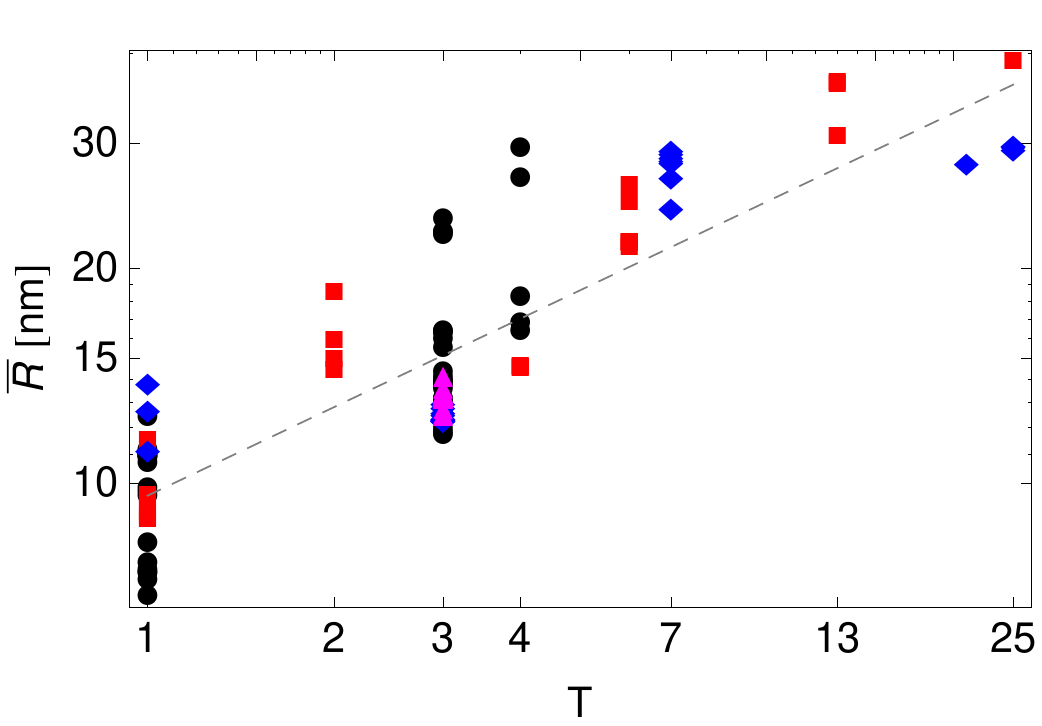}
\end{center}
\caption{Calculated mean radii of the capsids versus their triangulation numbers. The dashed line shows the least squares fit of the data to the 
function $\overline{R} = a\times T^{b}$, with $a=9.6\pm0.4$ nm, and $b=0.41\pm0.02$. The data consist of approximately 130 viruses, with circles denoting viruses with single-stranded genome, squares viruses with double-stranded genome, diamonds bacteriophages, and triangles $T=p3$ ssRNA viruses.}
\label{fig:mean_size_1}
\end{figure}

The sizes of viruses vary across a large interval of mean radii ($\sim$ 10-200 nm) and if there is something like a ``typical'' capsid (structural) protein, having some ``typical'' size, the larger viruses should contain more copies of it. This means that larger viruses should have a larger (triangulation) $T$-number~\footnote{There are $60T$ proteins in a virus of a given $T$-number.}, instead of being built of a smaller number of copies of a bigger protein and having a $T$-number comparable to smaller viruses. This proposition has been confirmed by Rossmann and Erickson~\cite{rossmann_erickson} on a dataset of viruses much smaller from the one we use, so it is of interest to reexamine this issue and update the analysis. 
To this end, we plot the mean radii of the capsids versus their triangulation numbers in Fig.~\ref{fig:mean_size_1}.

If the idea of a viral protein of some prototypical size makes sense, then the average capsid radius should be distributed according to 
\begin{equation}
\label{eq:roft}
\overline{R} = \left ( \frac{60 A_p}{4\pi} \right)^{1/2} T^{1/2},
\end{equation}
which was obtained simply from equating the area of an approximately spherical virus, $4\pi\overline{R}^2$, with the total area of assembled capsid proteins, $60 T A_p$, where $A_p$ is the average area per protein. The fit of the form
\begin{equation}
\overline{R}(T)=a\times T^b
\end{equation}
that we performed on the data shows that the idea indeed makes sense, as $b=0.41\pm0.02$ is quite close to the expected 0.5. The average area of a prototypical capsid protein is obtained from the coefficient $a$ and is found to be $A_p \approx 20\ \mathrm{nm}^2$. A similar result was obtained by Rossmann and Erickson~\cite{rossmann_erickson}. They find that the ratio of the virus radius and a square root of its $T$-number is approximately conserved, $R/\sqrt{T} \sim 10$ nm, which is in very good agreement with our results.

Here we mention that the choice of the viruses included in the fit can have a certain effect on the end results, as already noted elsewhere~\cite{Ting2011}. In Table~\ref{tab:fit} we compile the results of the fit for various subsets of the viruses included in our analysis. If we perform the fit on viruses with ss genome or with ds genome only, we obtain similar results as when the entire database is included. Bacteriophages, on the other hand, do not conform to the relationship in Eq.~(\ref{eq:roft}) that well, and excluding them from the fit gives a slope of almost $T^{1/2}$.

\begin{table}[!htb]
\begin{center}
\caption{Fitting coefficients for the dependence of capsid mean radius on its triangulation number, $\overline{R}(T)=a\times T^b$. The results of the fit are shown for the entire dataset, and for subsets selected on the basis of the type of the viral genome contained in the capsid.}
\begin{tabular}{|c|c|c|}
\hline
{\bf dataset} & a [nm] & b [\ ] \\
\hline
entire & $9.6\pm0.4$ & $0.41\pm0.02$\\
\hline
ss only & $9.1\pm0.6$ & $0.43\pm0.06$\\
\hline
ds only & $10.1\pm0.7$ & $0.45\pm0.03$\\
\hline
phage only & $11\pm1$ & $0.33\pm0.05$\\
\hline
ss and ds only & $8.75\pm0.3$ & $0.49\pm0.02$\\
\hline
\end{tabular}
\label{tab:fit}
\end{center}
\end{table}

We next investigate the average thickness of capsid proteins. In order to define the average capsid protein thickness we have analyzed the radial mass distribution of each capsid, which peaks around the capsid mean radius $\overline{R}$. We have defined the average capsid thickness $\delta$ as the full-width-at-half-maximum (FWHM) of this distribution (see Ref.~\cite{electro_jobp} for details). Note that our procedure implicitly assumes that the capsids are fairly spherical (as they in fact are for the present purpose). The width may be artificially enlarged due to the nonsphericity of the capsid, but this effect yields corrections of the magnitude which cannot falsify the similarities of thicknesses among different capsid proteins that we obtain.

\begin{figure}[!htb]
\begin{center}
\includegraphics[width=8.5cm]{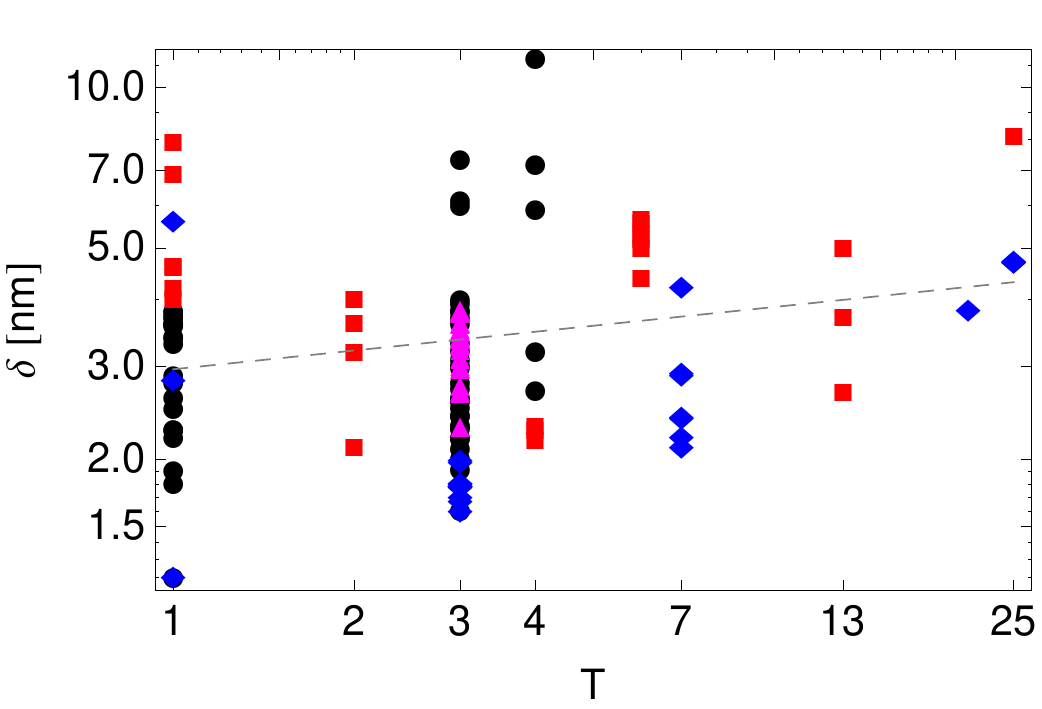}
\includegraphics[width=8.5cm]{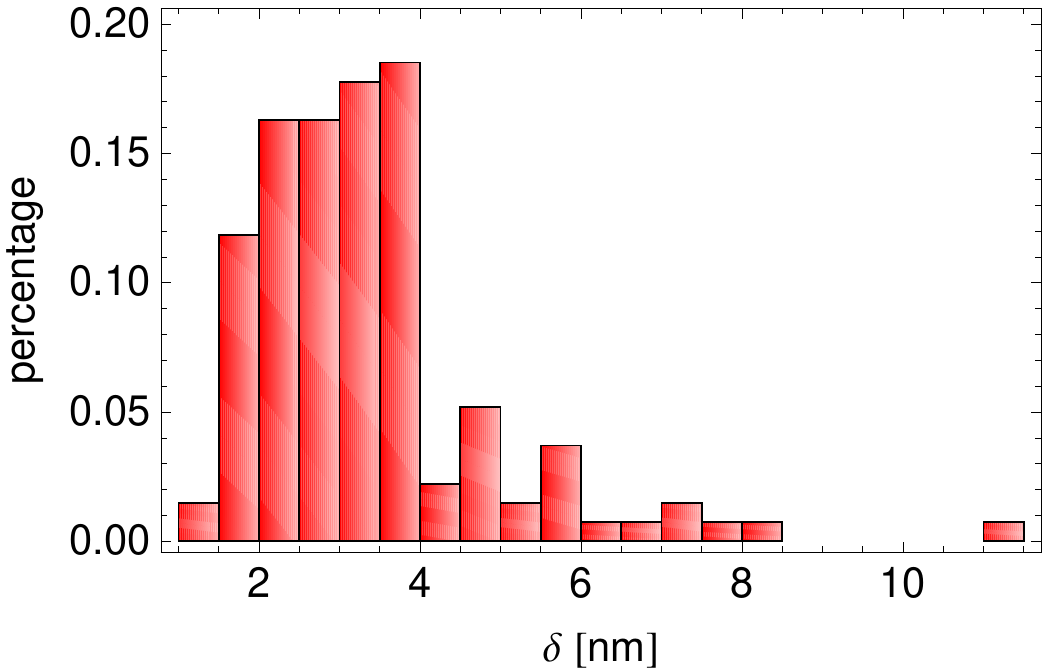}
\end{center}
\caption{ Distribution of capsid thicknesses. Top panel: Mean thicknesses of the capsids versus their triangulation numbers. The dashed line shows the least squares fit of the data of the form $\delta = a\times T^{b}$, with $a=2.9\pm0.2$ nm, and $b=0.12\pm0.05$. The symbols have the same meaning as in Fig.~\ref{fig:mean_size_1}. Bottom panel: Probability distribution of capsid thicknesses. Note that the about two thirds of the capsids in the set have thicknesses between 2 nm and 4 nm.}
\label{fig:mean_thickness}
\end{figure}

Our results for average capsid protein thicknesses are shown in Fig.~\ref{fig:mean_thickness}. The fit of the form $\delta (T)=a\times T^b$ gives a small positive exponent for $T$ dependence with $a=2.9\pm0.2$ nm and $b=0.12\pm0.05$. We thus find that the thicknesses of capsid proteins are almost constant and typically in the range of 2-4 nm, with thicker proteins being relatively rare, and intriguingly almost equally distributed in viruses with ss and ds genomes.

Some of the thicker viruses fall into the $T=1$ ds genome category, and a closer inspection suggests more pronounced protrusions around five-fold axes than in the $T=1$ viruses with the ss genome, which might explain the increased thickness. Some of the other outliers are two caliciviruses, Sindbis virus, Semliki Forest virus, and Nudaurelia capensis $\omega$ virus; no obvious characteristic is shared among these viruses.

\begin{figure}[!htb]
\begin{center}
\includegraphics[width=8.5cm]{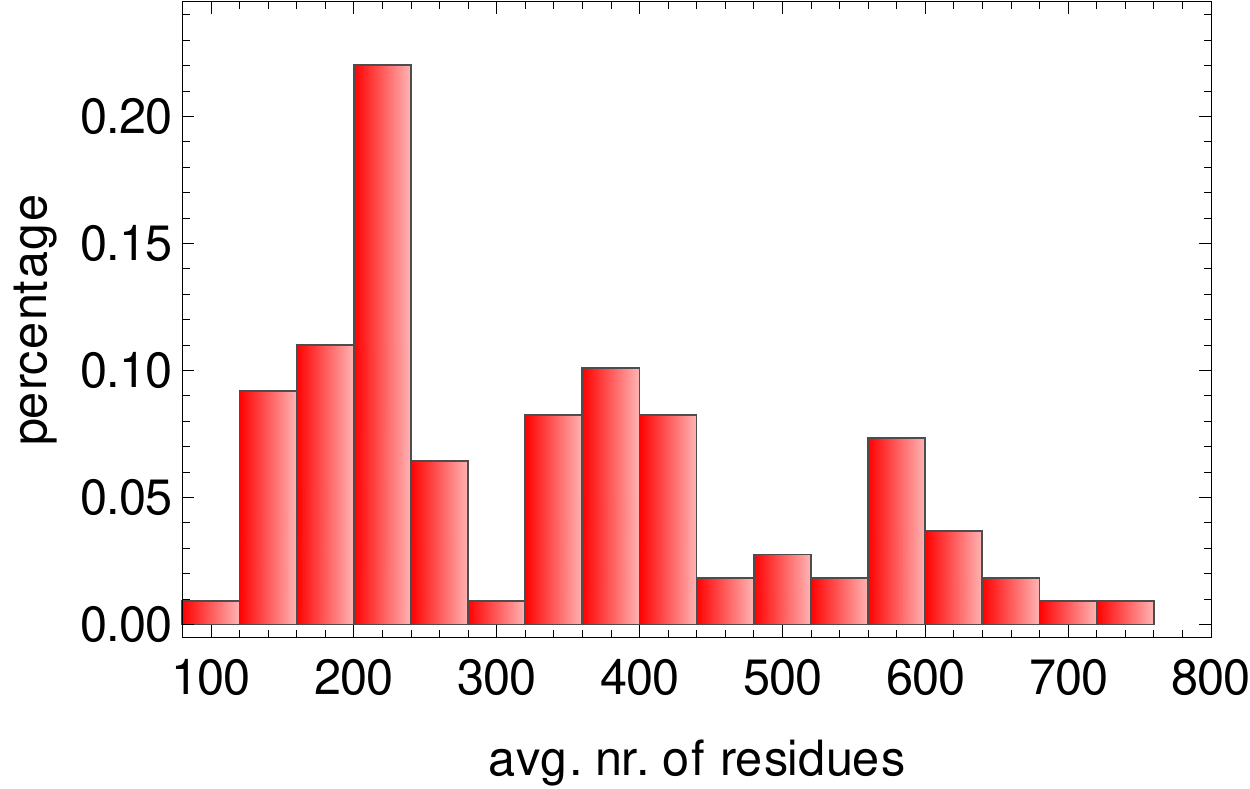}
\end{center}
\caption{Probability distribution of the average number of residues in capsid proteins. The average number of amino acids is obtained by averaging over $T$-number of constituent proteins in a capsid. There is a prominent peak at the average number of residues around $\simeq 200$ residues, and a smaller one around the average number of $\simeq 390$ residues.}
\label{fig:mass}
\end{figure}

Taken together, our analysis shows that the concept of a typical capsid protein is indeed viable -- the typical capsid protein is prism-like, about 3 nm thick and has an average diameter of 5 nm. The molecular weight pertaining to these protein dimensions can be estimated well by $\overline{m}_{prot}\simeq 2.7 \times 10^4\ \mathrm{amu}$, and the average number of amino acid residues peaks at $\simeq 200$ residues, with an additional smaller peak at $\simeq 390$ residues (Fig.~\ref{fig:mass}). Comparing this with the distribution of the most common sizes in a non-redundant representative set of protein chains (as given by e.g. PDBselect~\cite{PDBSELECT}) which has a peak at about $\simeq 140$ residues, one can observe that there are some viral coat proteins which are somewhat larger than the average. What interests us next are the elastic properties of the two-dimensional sheet made of such prototype proteins.

\subsection{Elasticity of Infinitely Thin Icosa(delta)hedral Shells: F\"{o}ppl-von K\'arm\'an Number}

The virus capsid is a shell -- as we have shown above, it is a structure formed by a layer of protein material ``wrapped'' so as to make a closed structure of icosahedral symmetry. The fact that the symmetry of the shell is icosahedral restricts the possible shapes of the capsid. Depending on the {\em elastic} properties of the protein capsid, i.e. the energetics of inter-protein contacts, such a shell may be more sphere-like, or more icosahedron-like. The sphere and the icosahedron are in fact two limiting shapes of the infinitely thin, continuum elastic icosahedrally symmetric shells with all the other allowed cases lying somewhere in between the two extremes. This opens the possibility to study the design and conserved elastic features of the virus proteins by analyzing the details of the shape of the icosahedral shell (capsid) they form.

The details of the elastic theory of icosa(delta)hedral shells without the spontaneous curvature have been elaborated in Refs.~\cite{LMN} and~\cite{Siber_LMN}. We shall present here only the most important issues relevant to our aim. It has been shown that the shape of a continuum icosahedral shell depends on {\em a single parameter}. This parameter is termed the F\"{o}ppl-von K\'arm\'an number ($\gamma$; FvK) in Ref.~\cite{LMN}, and it depends on the elastic properties of (formally infinitely thin) two-dimensional sheets made of virus proteins. There are {\em two} elastic parameters characterizing the elastic response of such sheets: the two-dimensional Young's modulus $Y$ which specifies the response of the sheet to stretching, and the bending rigidity $\kappa$ which quantifies the energetics of the sheet bending. The FvK number of a thin shell with mean radius $\overline{R}$ is a particular dimensionless combination of the elastic parameters 
of the shell and its radius, defined as
\begin{equation}
\gamma = \overline{R}^2\,\frac{Y}{\kappa}.
\end{equation}

\subsection{Fitting the Shapes: Determining Effective FvK Number of Real Viruses}\label{sec:fvk}

\begin{figure*}[!ht]
\begin{center}
\includegraphics[width=\columnwidth]{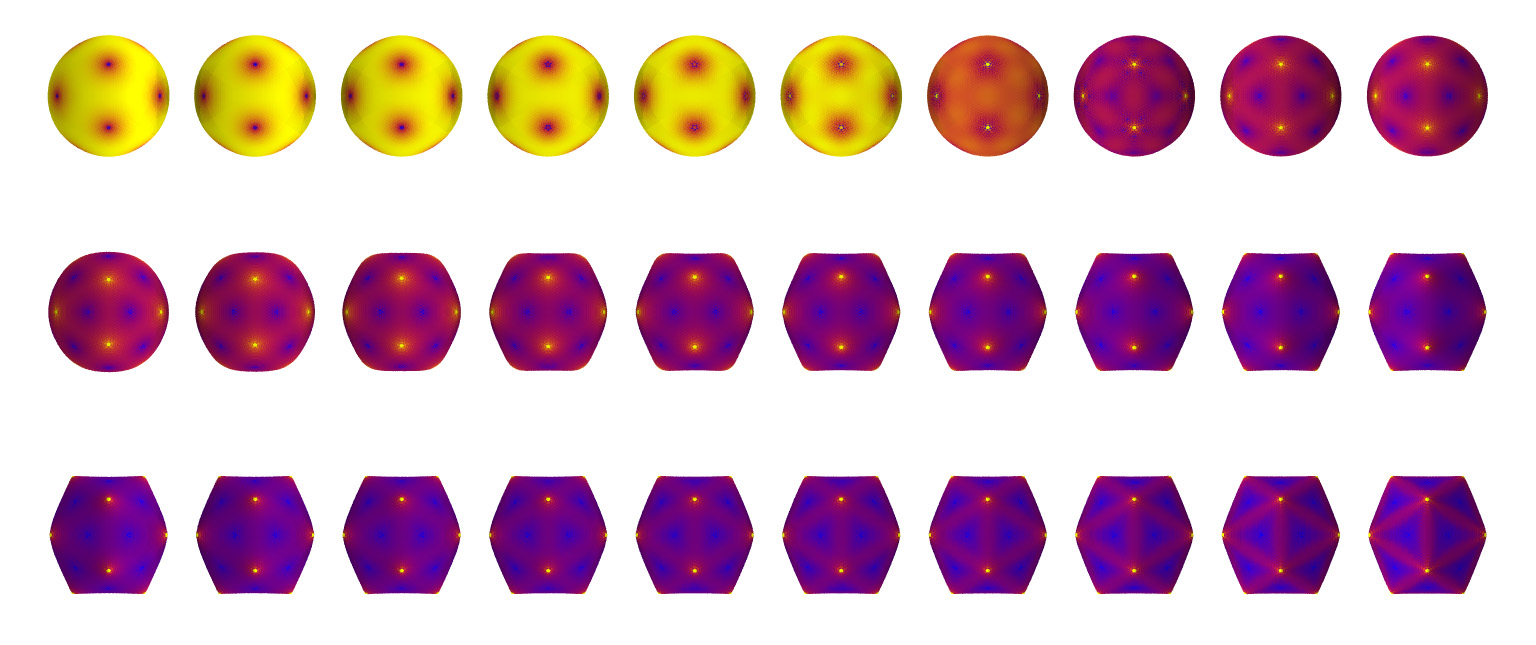}
\end{center}
\caption{Prototype shapes. The FvK number increases from left to right and from top to bottom. The range of FvK numbers goes from $\gamma=10^{-1}$ (almost a sphere) to $\gamma=10^7$ (almost an icosahedron), each shape in the sequence having FvK number larger from the previous one by a factor of $1.77$. The colors show the relative elastic energy contained in the parts of the shape surface, with yellow-colored regions having largest energy, magenta-colored ones having smallest energy, and blue-colored regions being in between (see Refs.~\cite{LMN} and~\cite{Siber_LMN} for details).}
\label{fig:fig_etalons}
\end{figure*}

We have approached the problem of extracting the elastic parameters of the proteins from the shape of the capsid by assigning an {\em effective} FvK number to a real capsid whose shape (i.e. the positions of atoms) has been determined experimentally. We have first generated sixty ``prototype'' shapes of different FvK numbers, i.e. the ideal shapes of the icosadeltahedral shells obtained for continuum shell material (formally infinite $T$-number; we have used $T=625$~\cite{Siber_LMN}). These shapes, 30 of which are shown in Fig. \ref{fig:fig_etalons}, were then compared to the real capsid and the one that fitted capsid the best was found. The FvK number of the best-fitting prototype shape was proclaimed to be the {\em effective FvK number of the virus}.

\begin{figure}[!ht]
\begin{center}
\includegraphics[width=8.5cm]{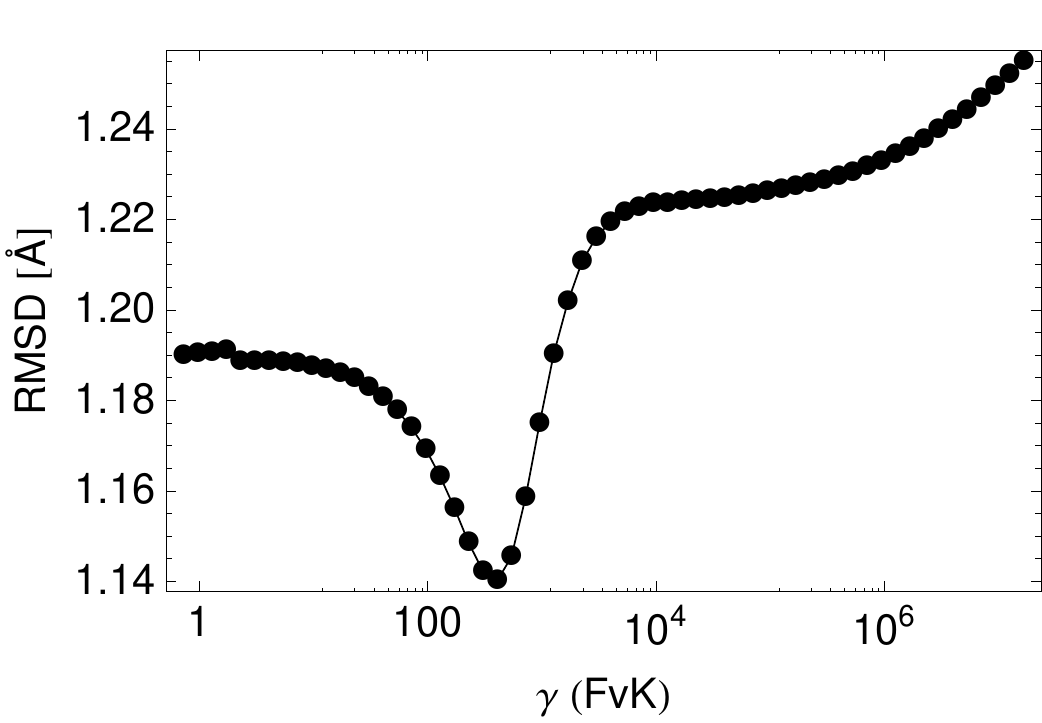}
\end{center}
\caption{Example determination of the effective FvK number of a virus. The virus in question is bacteriophage P22 procapsid (PDB ID 2xyy), and the effective FvK number assigned to it is the FvK number of the prototype shape with the minimum RMSD compared to the capsid. The procedure, detailed in the text, yields $\gamma=406$ in this case.}
\label{fig:phage}
\end{figure}

\begin{figure}[!ht]
\begin{center}
\includegraphics[width=8.5cm]{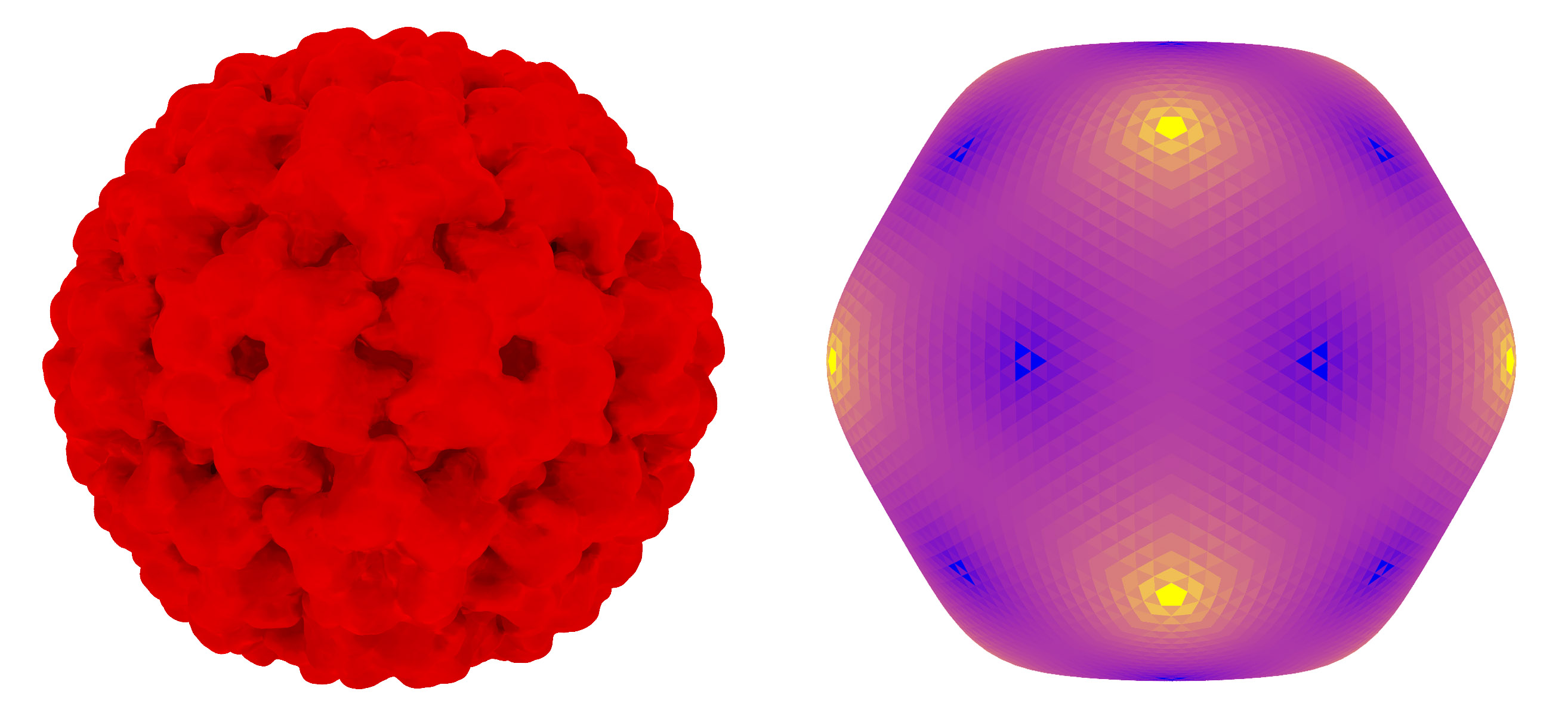}
\end{center}
\caption{Real capsid and the best-fitting prototype shape. Left: The experimentally determined shape of bacteriophage PM2 (PDB ID 2w0c), an icosahedral virus with triangulation number $T=p21$. The shape was constructed as a union of spheres of radius $8.4$ \AA, each one representing an amino acid of a virus. Right: The best-fit prototype shape with $\gamma = 720$.}
\label{fig:proto_fit}
\end{figure}

In order to conduct this scheme, one first needs to define the procedure for comparing shapes. We did this by calculating the root-mean-square deviation (RMSD) between a capsid and all 60 of the prototype shapes (details of the procedure are given in Sec.~\ref{sec:rmsd}). By doing this we obtained the $\mathrm{RMSD}(\gamma)$ dependence for our set of analyzed capsids. The minimum of this curve for each capsid is the effective FvK number associated with a given virus. An example curve is shown in Fig.~\ref{fig:phage} for the bacteriophage P22 procapsid (yielding a FvK number of $\gamma=406$), and Fig.~\ref{fig:proto_fit} shows a real virus shape of the bacteriophage PM2 and the corresponding best-fitting prototype with $\gamma=720$. Details of the procedure are given in the Analysis section.

A variant of the procedure described was most likely used in the analysis of L-A and HK97 viruses in Ref.~\cite{LMN} (the authors do not specify the details). Here we have further developed the procedure into a consistent fitting scheme and applied it to a large set of virus shapes as a tool for determination of elastic properties of capsid proteins. It is of interest to note here that for L-A and HK97 viruses we obtain values similar to those calculated in Ref.~\cite{LMN} (718 and 1270 vs. 547 and 1480 obtained in Ref.~\cite{LMN}).

The above procedure yields only {\em an estimate} of the elastic parameters of the capsids for four reasons: {\em (i)} The overall shape of the capsid may sometimes be importantly defined by the protein features not related to protein-protein interactions and the shell elasticity. Some capsids for example have spikes, troughs, and other peculiarities related to the special properties of the protein subunit. For such capsids, the procedure we propose may be less reliable. {\em (ii)} The theory was constructed for infinitely thin shells, so it should be applied with some reservation to viruses whose thickness is non-negligible with respect to their mean radius. {\em (iii)} The theory applies to sheets without spontaneous curvature. {\em (iv)} The theory applies to continuum shells, which means formally infinite $T$-numbers.
{\em (v)} The capsid mean radii are functions of external thermodynamic parameters, especially of the {\sl p}H value, and vary in quite different manners for different virus classes. {\em (vi)} Virus capsid is not really a closed shell but a shell with holes, where the holes are sometimes rather large; this can be the case either in the procapsid state or in the mature capsid.

\begin{figure}[!ht]
\begin{center}
\includegraphics[width=8.5cm]{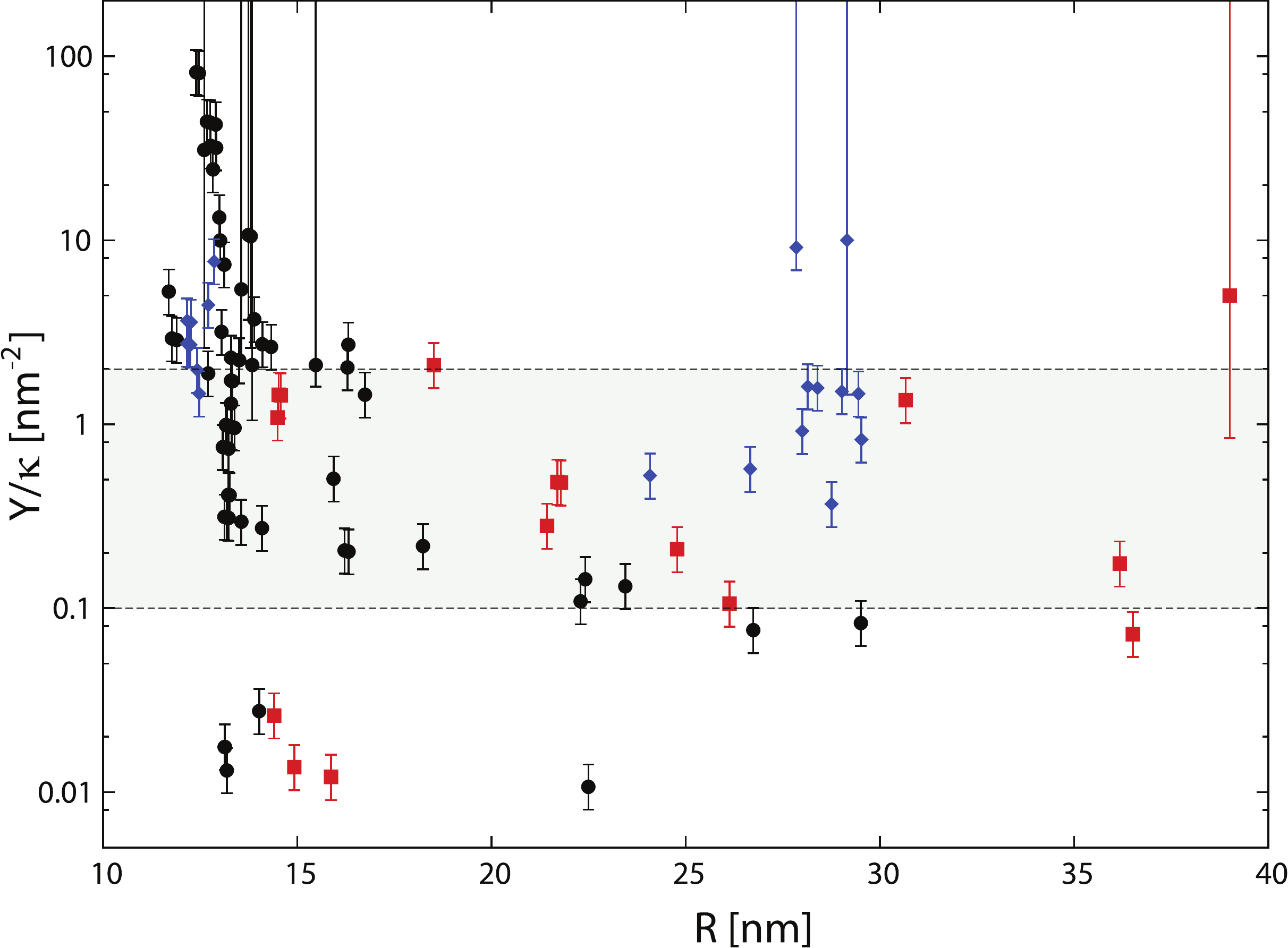}
\end{center}
\caption{Ratio of capsid elastic parameters $Y/\kappa$ compared to the capsid mean radius. The values of $Y/\kappa$ were obtained from best-fit FvK number, $\gamma=\overline{R}^2Y/\kappa$. The area between dashed lines shows the interval $[0.1,2]\ \mathrm{nm}^{-2}$, the values of the FvK numbers of the majority of $T>3$ capsids. Symbols have the same meaning as in Fig.~\ref{fig:mean_size_1}, with the exception of the $T=p3$ ssRNA viruses being shown in the same way as the rest of ssRNA viruses (circles). The size of the error bars depends on the sampling of FvK shapes -- it is the distance of the minimally deviating FvK number to the nearest FvK shape. The points with large error bars correspond to cases where the minimum is very shallow, so that only the left side of the uncertainty interval can be determined on the basis of the fit.}
\label{fig:yk}
\end{figure}

The point {\em (iv)} is particularly important for small viruses (see Ref.~\cite{aznar_small_T} for a discussion of the buckling of small-$T$ shells). To avoid false signals which may be obtained in the described way, we have decided to analyze only the capsids with $T>1$. For these capsids, the results of the fitting procedure are shown in Fig.~\ref{fig:yk}. For viruses with $T=2$ and $T=3$ a large spread of the ratio of the elastic parameters $Y/\kappa$ is observed, ranging from $0.01\ \mathrm{nm}^{-2}$ to almost $100\ \mathrm{nm}^{-2}$.

When we move to higher $T$-numbers -- viruses with $T>3$ -- the data show a reasonably converged distribution of $Y/\kappa$ values, with most of the values falling into the interval $[0.1, 2]\ \mathrm{nm}^{-2}$, indicated by the gray rectangle in Fig.~\ref{fig:yk}. The notable exception though is the point located at mean radius of $\overline{R}=27.85\ \mathrm{nm}$, having the value of the ratio of $Y / \kappa = 9.1\ \mathrm{nm}^{-2}$; this point belongs to PM2 bacteriophage. PM2 bacteriophage is a special virus in the sense that below the outer protein shell it contains a mixed lipid-protein membrane (i.e. it has a proteinaceous lipid core)~\cite{PM2}. However, so does the PRD1 bacteriophage~\cite{PRD1}, whose $Y / \kappa$ ratio is within the typical range of values (0.8-1.5 $\mathrm{nm}^{-2}$). The reason for the observed large $Y / \kappa$ ratio in PM2 is thus most likely due to its very pronounced spike proteins around the five-fold axes which may influence our shape analysis and push its effective FvK number towards larger values; the PRD1 bacteriophage, on the other hand, has no such prominent features. These two cases again accentuate the care required in interpretation of our results.

The reason for the wide-spread range of the $Y/\kappa$ values for $T=3$ viruses and a much narrower range of this ratio for viruses with higher $T$-numbers might be explained, at least qualitatively and conceptually, by a transition between a mostly geometric domain and a continuum domain as proposed in Ref.~\cite{Mannige2010}. In the former domain the geometry of the protein subunits influences the capsid morphology. After a certain capsid size or $T$-number the shapes are then largely described by continuum elasticity where FvK number is the relevant parameter. Due to our limited dataset it is difficult to say where exactly this transition occurs, but the Fig.~\ref{fig:yk} suggests that this happens somewhere around $\overline{R}=20\ \mathrm{nm}$. Intriguingly, a very similar finding was reported recently in Ref.~\cite{CLB_III}, where the authors report a strong correlation between $\gamma$ and a ``degree of buckling'' in $T=7$ capsids, and lack of such a correlation in $T=3$ capsids.

Bending rigidity is related to the energy required to change the angle between two capsid proteins in flat contact, while the Young's modulus measures how difficult it is to stretch the two capsid proteins in contact, keeping them flat. Having in mind all the intricacies of protein-protein interaction, it is not easy to see that the two quantities ought to be in any particular relation, so that they should be, in general, treated as completely independent parameters. The fact that the {\em ratio} of two independent elastic parameters falls in an interval of only order of magnitude wide points again to a possible conservation of this quantity.

The ratio $Y/\kappa$ can be obtained as a function of thickness of the shell in the case of an isotropic elastic material, which the protein shell is obviously not, as the proteins preferably bind only in 2D. Nevertheless, if we insist on approximating the proteins as isotropically interacting particles, we have that $Y / \kappa = 12 (1 - \nu^2) / \delta^2$, where $\delta$ is again the shell thickness and $\nu$ the Poisson ratio of the material. Interestingly, for $\delta \sim 3\ \mathrm{nm}$, and $\nu \sim 0.3$, which is typical for many materials, this gives $Y / \kappa \sim 1\ \mathrm{nm}^{-2}$, which fits nicely in the range we obtained (but also with the values found in Refs.~\cite{LMN} and~\cite{CLB_III}), suggesting that the elastic response is essentially fixed with conservation of $\delta$ (i.e. conservation of protein size we have already established). This is further corroborated by the examination of extremal values of $\delta$ and $Y / \kappa$ in our set. Sindbis virus (PDB ID 1ld4) has the largest $\delta$ in our dataset (11.27 nm), but also one of the smallest ratios of $Y / \kappa$, (0.08 nm$^{-2}$), consistent with a simple prediction  $Y / \kappa \propto \delta^{-2}$. However, when we constructed the quantity $Y \delta^{2} / \kappa$ we found that its spread is not significantly smaller than the spread of $Y / \kappa$ values, suggesting that this ratio contains more information than simply $\delta$.

\section{Summary and Conclusions} 

Using the statistical analysis we have shown that capsid proteins of all viruses are similarly sized, prism-like, about 3 nm thick, having an average diameter of 5 nm and an average molecular weight of $\overline{m}_{prot}\simeq 2.7 \times 10^4\ \mathrm{amu}$. This is by no means a trivial finding. In the early days of virus structure research~\cite{Morgan} it has been argued that viruses cannot code for large structural proteins as it would require long genomes, and the capsid thus must be assembled from many copies of a smaller protein. Although this sounds reasonable one should not forget that the quantity of information that can be stored in the capsid scales as $\overline{R}^3$, presuming the genome is uniformly distributed within the capsid, which is the case at least for bacteriophages~\cite{Catalano,Gelbart-PT}. Were all capsids made by assembling 60 protein units ($T=1$) of similar thickness, the quantity of information required for coding the capsid proteins would then scale as $\overline{R}^2$, so the percentage of information required for structural proteins would be vanishingly small for large enough viruses, i.e. large enough genomes. This, seemingly obvious statement is perhaps even more strengthened by the fact that there {\em are} huge viruses where the spatial constraints do not seem to be the critical issue~\footnote{The analysis of available space and restricted length of the genome is somewhat different in ssRNA viruses, where the ssRNA molecule is held within the capsid mostly by electrostatic interactions with the capsid proteins. In this case, the quantity of information that can be stored in the capsid scales as $\overline{R}^2$~\cite{PCCP}, so that the production of large proteins always requires the same percentage of the ssRNA information, irrespectively of the radius of the virus.}. Yet there are apparently no huge icosahedral viruses with small $T$-numbers, i.e. made of huge capsid proteins. The capsid protein size appears thus to be an evolutionary conserved feature.

An important question is whether the elasticity of the capsid is a property which is under evolutionary pressure and thus makes a difference for the functioning of a virus. There is at least one type of viruses for which we can be fairly certain that the answer is affirmative -- the dsDNA bacteriophages~\cite{Gelbart-PT}. The bacteriophages pack their dsDNA molecule in the preformed capsid tightly, building up an effective outward mechanical pressures on the capsid up to almost a hundred atmospheres. Such huge pressures would induce large displacements in elastically soft capsid material, resulting in eventual rupture of protein contacts formed by either through hydrophobic-van der Waals association or electrostatic complementarity. Thus, at least for these types of viruses the elastic properties of capsids should be under evolutionary pressure and should converge to some functional range.

This does not exclude the possibility of the conservation of  elastic properties in other types of viruses, where possibly different types of elastic constraints might be at work. The situation of RNA virus capsids is in fact exactly reversed, as the (small) force of the genome on the capsid creates an effective inward mechanical pressure, being of electrostatic bridging origin~\cite{PCCP}. The elastic constraints in this case would play a prominent role not so much in the context of structural rigidity of the capsid as in the whole self-assembly and maturation process.

Last but not least, in order to penetrate the cell viruses undergo often complicated and multi-stepped paths (i.e. receptor attachment, membrane wrapping, etc.) which include many interactions, eventually resulting in effective mechanical forces on the capsid. The architecture of the capsid and mechanical properties of its building blocks must be well suited to successfully complete this, maybe the most fragile, part of the viral life-cycle.

By combining the statistical analysis with the theory of elasticity we have analyzed the elastic properties of the virus capsids. Our results suggest a reasonable convergence of the elastic properties of the viruses we inspected. To fully evaluate the power of  the methods we proposed to discriminate among different virus architectures and possibly even lineages, a significantly larger virus dataset would need to be analyzed which is at present not possible.

\begin{acknowledgements}
We thank Alberto Vianelli for informing us about Ref.~\cite{rossmann_erickson}. We also thank the two anonymous reviewers for their suggestions on improving the manuscript. A.L.B. acknowledges the support from the Slovene Agency for Research and Development under the young researcher grant. A.\v{S}. acknowledges support from the Ministry of Science, Education, and Sports of Republic of Croatia (Grant No. 035-0352828-2837). R.P. acknowledges the support from the Slovene Agency for Research and Development (Grant No. P1-0055 and J1-4297).
\end{acknowledgements}


\end{document}